# Rheopexy and tunable yield stress of carbon black suspensions

Guillaume Ovarlez, Laurent Tocquer, François Bertrand and Philippe Coussot

*Université Paris-Est, Laboratoire Navier (UMR 8205), CNRS, ENPC, IFSTTAR, F-77420 Marne-la-Vallée*

**Abstract:** We show that besides simple or thixotropic yield stress fluids there exists a third class of yield stress fluids. This is illustrated through the rheological behavior of a carbon black suspension, which is shown to exhibit a viscosity bifurcation effect around a critical stress along with rheopectic trends, i.e., after a preshear at a given stress the fluid tends to accelerate when it is submitted to a lower stress. Viscosity bifurcation displays here original features: the yield stress and the critical shear rate depend on the previous flow history. The most spectacular property due to these specificities is that the material structure can be adjusted at will through an appropriate flow history. In particular it is possible to tune the material yield stress to arbitrary low values. A simple model assuming that the stress is the sum of one component due to structure deformation and one component due to hydrodynamic interactions predicts all rheological trends observed and appears to well represent quantitatively the data.

## 1. Introduction

Materials such as dense suspensions, colloidal gels, microgel suspensions, concentrated emulsions or foams, are yield stress fluids. They can flow only when the applied stress is higher than a critical value $\tau_y$; below $\tau_y$, they have a solid viscoelastic behavior which results from the jamming of their structure[1-5]. Two different classes of yield stress fluids have been recently distinguished based on the characteristics of their flow behavior at the approach of the jamming transition[6-9].

Some of these materials see their flows continuously slow down and become infinitely slow when the applied stress approaches the yield stress $\tau_y$. In other words, their apparent viscosity diverges when $\tau \to \tau_y^+$, which can also be formalized as $\dot{\gamma} \xrightarrow[\tau \to \tau_y^+]{} 0$, as in classical (Bingham, Herschel-Bulkley) steady-state constitutive laws $\tau(\dot{\gamma})$ of yield stress fluids. The value $\tau_y$ below which their flows stop ('dynamic' yield stress) is a well-defined unique property of these materials; the critical shear stress value at flow start-up ('static' yield stress) is usually close to it. It has been suggested to call such materials "simple yield stress fluids". From experimental observations, they seem to be mostly composed of nonthixotropic dense suspensions of soft particles (microgel suspensions, concentrated emulsions, foams)[10].

By contrast, some other materials experience an abrupt transition from flow to rest at the jamming transition. When the applied stress $\tau$ is above $\tau_y$, their steady-state apparent viscosity $\eta$ takes a finite value lower than a critical value $\eta_c$; when $\tau$ is lowered below $\tau_y$, $\eta$ suddenly becomes infinite. Such a trend led to call this phenomenon "viscosity bifurcation"[11]. Consequently, these materials cannot experience slow steady flows: the steady-state shear rate characterizing their flows is always larger than a critical value $\dot{\gamma}_c = \tau_y / \eta_c$, which can also be formalized as $\dot{\gamma} \xrightarrow[\tau \to \tau_y^+]{} \dot{\gamma}_c$. When a flow is imposed at low macroscopic shear rate, this leads to shear banding, i.e., to the separation of the material into two regions: a non-flowing region, and a region flowing at $\dot{\gamma}_c$ [12-14]. The liquid/solid transition in these materials is thus characterized by two independent intrinsic material properties: the critical values $\tau_y$ and $\dot{\gamma}_c$ below which flow stops, which take well-defined unique values. Note however that the critical values at flow start-up depend on the flow history and on the resting time. Viscosity bifurcation has been observed in all studied thixotropic yield stress fluids[7,13,15-18] and in density-mismatched noncolloidal suspensions[19]. From a physical point of view, it seems to result from shear/structure coupling, through a competition between a restructuring process (e.g., aggregation in colloidal gels) and shear-induced destructuration. Several models based on the idea that the material properties depend on its structure and that the structure depends on shear history[13,20,21] have indeed predicted viscosity bifurcation.

Although they are much less often encountered, shear-history-dependent materials of other kind exist, namely rheopectic materials (also called anti-thixotropic materials). Starting from a steady flow at a given shear stress or shear rate, these materials see their apparent viscosity increase in time when the applied stress of shear rate is increased. From a structural point of view, this means that they get structured under shear. When the applied stress or shear rate is lowered, their apparent viscosity decreases in time, i.e., they get destructured. As far as we know, the possible yielding behavior of rheopectic materials has not been investigated in the literature.

From the basic features of this behavior, it can first be wondered whether rheopectic yield stress fluids exist or not and what are the physical ingredients that may explain their existence, since they should get less and less jammed as the applied stress is lowered. The next question that arises is the situation of such materials with regards to the two classes of yield stress fluids that has been identified up to now (see above).

In this paper, we study the rheological behavior of a material which has been shown to be rheopectic – namely, a suspension of carbon black particles – with a special focus on the characteristics of the liquid/solid transition. We show that, starting from any flowing state, although their steady-state flow curve displays no yield stress, these materials show viscosity bifurcation. However, their behavior contrasts with that of thixotropic yield stress fluids, since their dynamic yield stress $\tau_y$ and their critical shear rate $\dot{\gamma}_c$ depend on the previous flow history. In addition, viscosity bifurcation in these suspensions is not associated with shear banding. These materials thus constitute a new original class of yield stress fluids.

We then show that it is possible to predict all their original characteristics as a function of flow history with a model based on very simple arguments: the shear stress is assumed to be the sum of a flow rate-independent term related to the current structure state and a flow rate-dependent term related to hydrodynamic effects. The first term evolves as a function of flow history according to a function of the material. This approach provides some basic understanding of the physical origin of the behaviors of these suspensions.

## 2. Materials and methods

In the literature, materials that have several times been observed to behave as rheopectic materials are carbon black suspensions[22-24]. In these materials, small elementary particles ($\approx 30$nm) are irreversibly flocculated into large fractal particles ($\approx 0.5\,\mu$m) which interact through low attractive Van der Waals forces and may form a colloidal gel. This multiscale structure plays a major role in their behavior as discussed by Osuji et al.[25] As in any colloidal gel, aggregation is promoted at low shear, and aggregates are broken at high shear rates. When aggregated, the fractal particles have the possibility to arrange into dense aggregates of interpenetrated particles, which have a density larger than the individual fractal particles. Suspensions of aggregates thus have a lower viscosity than the suspensions of fractal particles, in contrast with gels made of non-fractal particles. This leads to a rheopectic behavior, whereas gels made of non-fractal particles usually have a thixotropic behavior. The materials formed under shear have been shown to have a solid viscoelastic behavior[23-26], which supports the formation of a loose colloidal gel under shear. These materials should thus behave as yield stress fluids, but the features of this behavior have not been studied in detail in the literature. The local flow properties at flow start-up of a carbon black suspension have been studied by Gibaud et al.[27]; they have shown that flow starts inhomogeneously, following a complex scenario observed in other yield stress fluids[28,29]. However, the behavior at flow start-up of yield stress fluids is known to be much more complex than that at flow stoppage[10,29], and it does not provide the same information.

Here our aim is to clarify the behavior of these suspensions with regards to the usual features of yield stress fluids. In particular we will focus on the steady-state flow behavior of the material and on the possible viscosity bifurcation phenomenon at the jamming transition, i.e., on the stability of flows. This should be investigated by first imposing a homogeneous flow of the material in its liquid regime and then decreasing the applied load[10]. In the following, we will thus study the response of the material to steps in shear stress, for various flow histories. We will also use MRI data to validate some deductions from macroscopic data.

Carbon black suspensions are prepared following Gibaud et al.[27] and Trappe et al.[30]. Carbon black particles (Cabot Vulcan XC72R, density 1.8) are dispersed at a weight concentration of 6% in a light mineral oil (Sigma, density 0.838, viscosity 20 mPa.s) with a standard mixer. The sample is further sonicated for 1 hour, and then left at rest for 1 day. The entire batch is vigorously stirred again before any rheological study. With this procedure, experiments performed on a same batch were reproducible for several months.

Rheological studies are performed with a Malvern Kinexus rheometer equipped with a thin gap Couette geometry (inner cylinder radius 12.5 mm; outer cylinder radius: 13.75 mm) with serrated surfaces. The gap size (1.25 mm) should be sufficiently large to avoid the formation of vorticity-aligned flocs[26,31], i.e., to ensure that we study the material bulk properties. Such structures, of diameter equal to the gap size, may indeed appear only at shear rates lower than $\simeq 0.1 \text{ s}^{-1}$ for gaps of size larger than 1 mm[31].

Other experiments are performed within a wide gap Couette geometry (inner cylinder radius: 4.1 cm; outer cylinder radius: 6 cm; height $H$ of the inner cylinder: 11 cm) inserted in a MRI display described in detail previously[32,33], in order to obtain the local flow properties of the materials and to validate the results form macroscopic experiments. Azimuthal velocity profiles $V(R,t)$ are measured as a function of the radial position $R$ in the gap and of time $t$ using MRI techniques described previously[32,33], for various constant rotational velocities $\Omega$ ranging between 1 and 100 rpm; the torque $T$ applied on the inner cylinder is also measured. The profiles are averages over 8 s; the radial resolution is 250 $\mu$m. We recall that the local shear rate

$$\dot{\gamma}(R,t) = V(R,t)/R - \mathrm{d}V(R,t)/\mathrm{d}R \qquad (1)$$

can be deduced from the velocity profiles and that the local shear stress

$$\tau(R) = T/(2\pi R^2 H) \tag{2}$$

can be computed everywhere from the momentum equation.

## 3. Experimental results

### *Creep tests*

In order to study the liquid/solid transition in the carbon black suspension, we perform creep tests with the following procedure. We first preshear the material at high imposed shear stress $\tau_0$ until a steady state flow is reached: this defines the initial state of the material. Then we abruptly decrease the applied shear stress to a new value $\tau_i$ and we study the time evolution of the shear rate $\dot{\gamma}(t)$. The same experiment is performed several times, for various $\tau_i$ values.

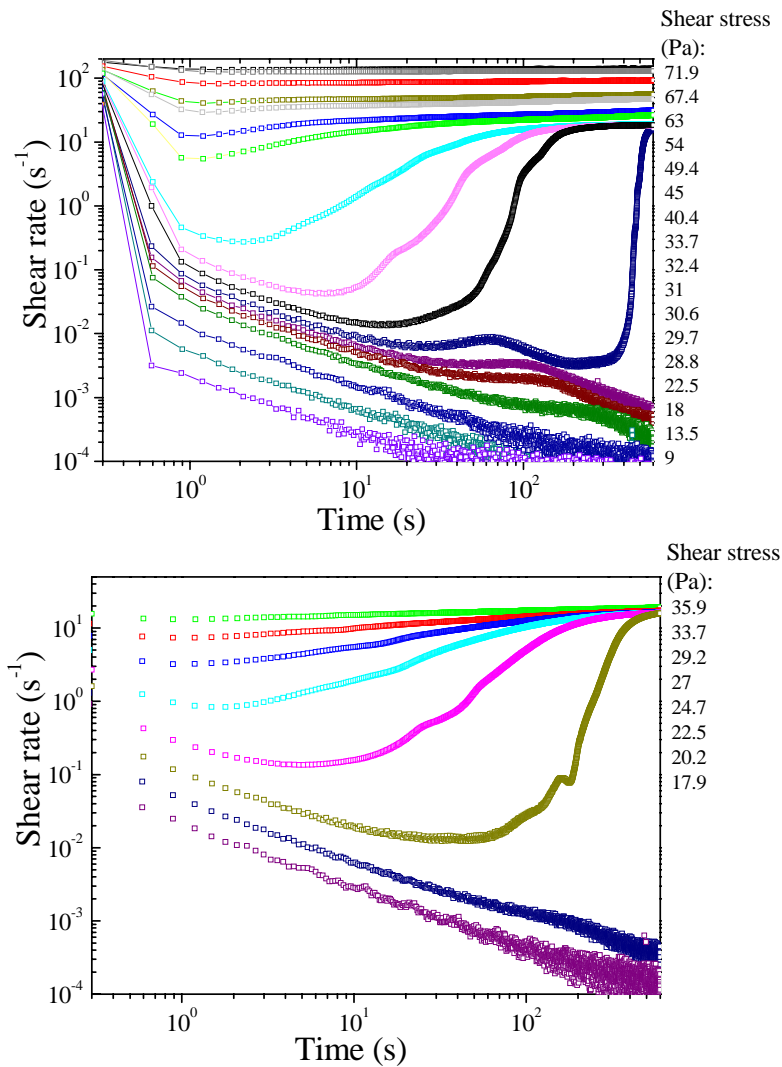

***Fig. 1****: Shear rate vs. time for creep tests performed at different stress values $\tau_i$ after a preshear at $\tau_0 = 80.9$ Pa (top) and a preshear at $\tau_0 = 35.9$ Pa (bottom). The stress values are indicated on the right of the panels, in the same order as the curves.*

In Fig. 1a, we show the response to such creep tests of a carbon black suspension presheared at a stress $\tau_0 = 80.9$ Pa, for various subsequent stress values $\tau_i$. The first important observation from these creep tests is that the material clearly has a **yield stress**:

- for stresses smaller than $\tau_y \approx 30.5$ Pa, the material progressively stops flowing, i.e. ,the shear rate continuously decreases to very low values and the deformation tends to saturate, which means that the material is in its solid regime;

- for stress values above $\tau_y$, after the very first times (say, beyond 1 s) the flow accelerates in time until the shear rate reaches a steady-state value; the material thus flows steadily: it has reached a liquid regime.

The second important observation is that this material is **rheopectic**. Indeed, when we impose a lower shear stress after a steady flow at a high shear stress, the apparent viscosity of the material increases. This suggests that the fluid reaches a more jammed state of structure when it flows at a high shear stress than at a small shear stress.

The third important observation is that a spectacular **viscosity bifurcation** occurs around the jamming transition: steady-state flows, even for stresses very close to $\tau_y$, are all characterized by large shear rate values, higher than a critical value $\dot{\gamma}_c \approx 18 \text{ s}^{-1}$. At first glance, this rheopectic suspension thus seems to share common features with thixotropic materials at the liquid/solid transition.

Such observations are reproducible for the same level of preshear stress and qualitatively similar for other levels of preshear stress. However, when looking at the detailed response to creep tests performed after another preshear stress level an original phenomenon appears. Let us for example consider a series of creep tests performed after applying a preshear stress $\tau_0 = 35.9$ Pa significantly smaller than in the first series of experiments, but still larger than the apparent yield stress found in this series.

The response of the carbon black suspension to these experiments is depicted in Fig. 1b. Again, we observe viscosity bifurcation. However, surprisingly, this phenomenon is here characterized by two new, lower, critical values $\tau_y \approx 22$ Pa and $\dot{\gamma}_c \approx 16 \text{ s}^{-1}$. This implies in particular that, depending on flow history, the material may stop flowing or not for a given stress level. E.g., in Fig. 1, the material stops flowing when a shear stress equal to 22 or 29 Pa is applied just after a 80.9 Pa preshear, whereas it flows and reaches a steady shear rate value $\approx 16 \text{ s}^{-1}$ when the same stresses are applied after a 35.9 Pa preshear.

If we now perform another series of creep experiments on a material presheared for example at $\tau_0 = 24.7$ Pa a viscosity bifurcation is again observed (data not shown), characterized by two other critical values $\tau_y \approx 14$ Pa and $\dot{\gamma}_c \approx 12 \text{ s}^{-1}$. Such an experiment leading to the same qualitative trends may be repeated indefinitely at smaller preshear stress levels. Such a material finally has a yield stress $\tau_y$ and a critical shear rate $\dot{\gamma}_c$ which depend on the flow history, which contrasts with other kinds of yield stress fluids. More precisely $\tau_y$ and $\dot{\gamma}_c$ increase with the intensity of the preshear flow. In order to characterize more extensively the dependence of the material yield stress on shear history, we have made complimentary experiments where the yield stress is measured with the procedure described below (see Fig. 4b) on the material presheared at various stress values. The results are depicted in Fig. 2.

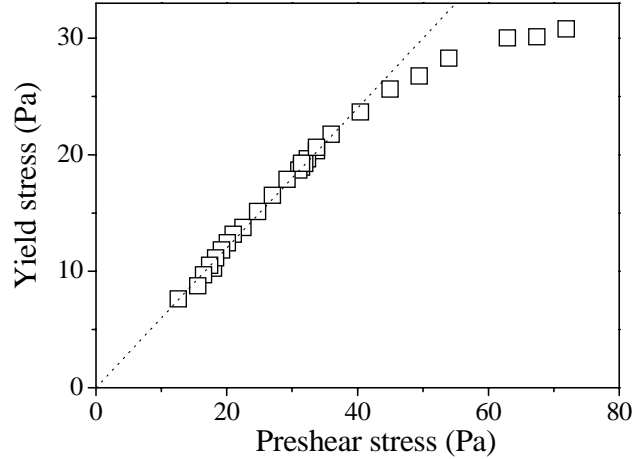

*Fig. 2: Yield stress of carbon black suspensions prepared at various preshear stress values (see text). The dotted line is a straight line of slope 0.6 (guide for the eye)*

Note that if we now start again a creep test series with a higher stress level, for example at $\tau_0 = 80.9$ Pa, we will get exactly the same results as for the initial tests. This means that the evolutions of the state of structure are perfectly reversible.

*Steady-state flow curve*

We now focus on the steady-state flow behavior of the material. From the above experiments, we see that the material may be able to flow at a given applied shear stress provided the shear stress applied just before is not too high. In order to investigate the ability of the material to flow steadily at low applied stresses, we thus used the following procedure. We applied a decreasing series of constant shear stresses $\tau_j$. At step $j+1$ the shear stress value is set as $\tau_{j+1} = 0.95 \times \tau_j$. Each stress is applied during 15 min. The stress applied at the first step is $\tau_1 = 80.9$ Pa; the stress applied at the last step is $\tau_{96} = 0.6$ Pa; the experiment lasts 24 h. The response to this succession of creep tests is shown in Fig. 3a; the steady-state flow curve extracted from these experiments is shown in Fig. 3b.

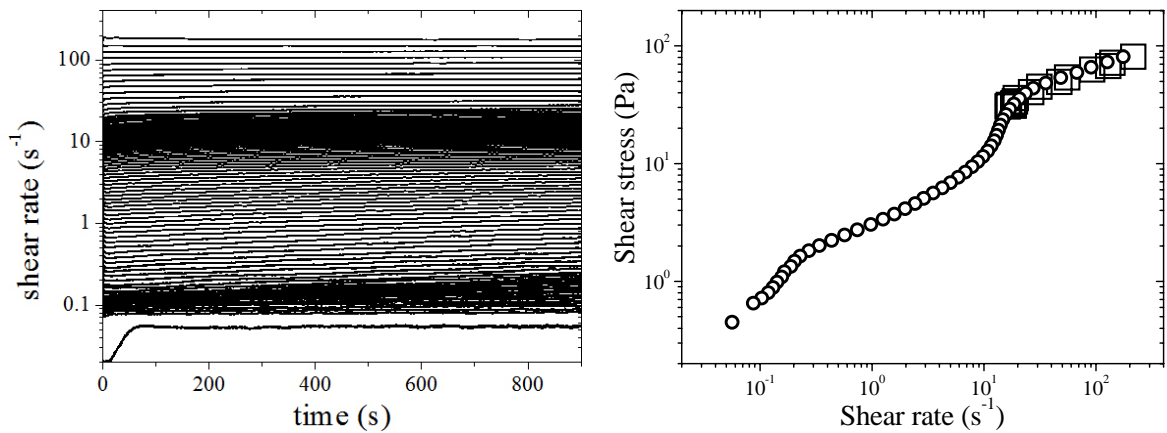

*Fig. 3: Left: shear rate vs. time for a succession of decreasing applied stresses $\tau_j$ (with $\tau_{j+1} = 0.95 * \tau_j$). Right: steady state flow curve extracted from Fig. 3a (crosses) and from Fig. 1a (empty squares).*

We now observe that the steady-state flow curve is not that of a yield stress material: there is no plateau at all at low shear rates, the material seems to have the ability to flow at any low applied stress level (as small as 0.6 Pa); moreover, it seems that a flow can occur at any low value of the shear rate (as small as 0.1 s$^{-1}$). This strongly contrasts with the behavior observed in the creep experiments (see above) in which no flow could be observed below a critical stress and a critical shear rate. The difference between the responses to the two experiments is further illustrated by the comparison of the steady-state shear stress vs. shear rate values extracted from them: in Fig. 3b, we see that only a small part of the steady-state flow curve is accessed when all creep tests are performed just after a preshear at a same given value $\tau_0$ (in Fig. 3b, $\tau_0 = 80.9$ Pa).

It thus seems that the material prepared by a preshear at a stress $\tau_0$ is in a state that is frozen if the applied stress is abruptly decreased below a given yield stress value $\tau_y$ (which is a function of $\tau_0$) whereas the material state can slowly evolve towards a less and less jammed state when the stress is slowly decreased.

*Characterization of the "intrinsic" behavior of the presheared material*

To better characterize the state of the material obtained by a preshear at a stress $\tau_0$, we now try to describe its **intrinsic** flow behavior, i.e. the constitutive law $\tau(\dot{\gamma})$ that would characterize its flows if the material could be kept in the same state at other values of $\tau$. In principle, this may be done by collecting the instantaneous shear rate values $\dot{\gamma}(\tau_i, t = 0^+)$ reached after the step changes in shear stress, from $\tau_0$ to the various studied $\tau_i$ values (such an analysis is classical for thixotropic materials**34-38**). However, this task is experimentally difficult to achieve as the measured short time response may involve both material response and inertia effects. As the material is rheopectic and gets destructured under shear when the stress is decreased, it is likely that the possible initial decrease in shear rate observed over short time is mainly related to inertia effects. As a consequence we have chosen to associate to each stress value $\tau_i$ the minimum shear rate value reached during a creep test, which should be characteristic of the most structured state visited during this flow and should thus be as characteristic as possible of the presheared material. In most cases, this minimum is obtained within a few seconds but obviously our choice is somewhat questionable for stresses close to the yield stress where the minimum shear rate is reached after a few tens of seconds. For stresses below the yield stress there is a progressive decrease of the shear rate and the deformation tends to saturate, which indicates that we are in the solid regime, the corresponding apparent shear rate cannot be used in the flow curve associated with the liquid regime. The results of this analysis, for three different values of the preshear stress $\tau_0$, are depicted in Fig. 4a.

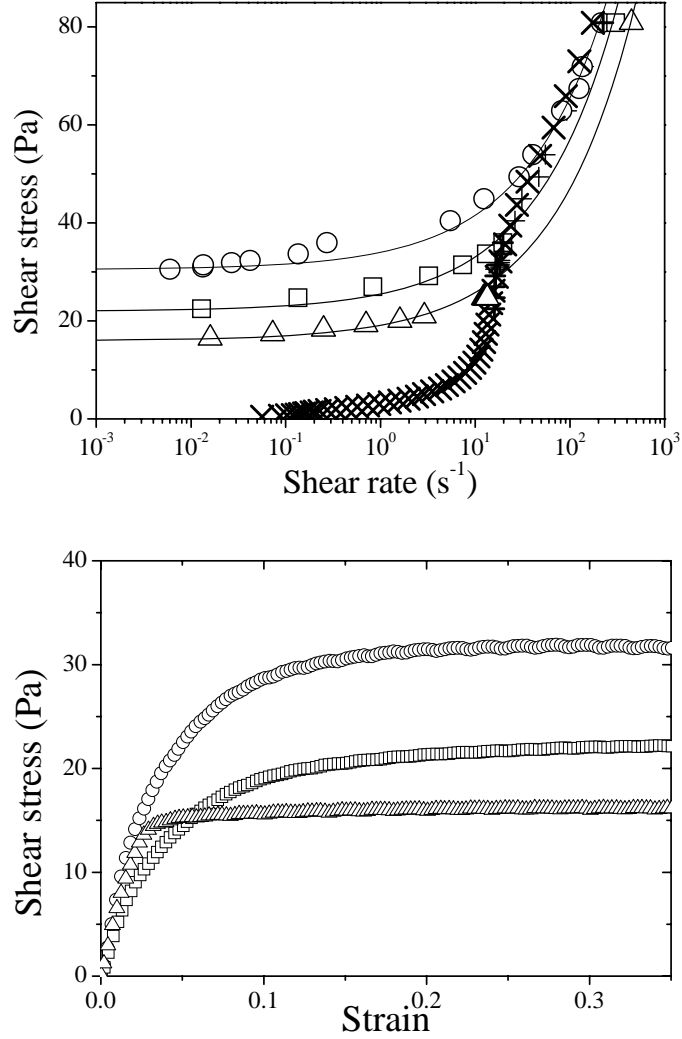

*Fig. 4: Top: "intrinsic" flow curves (open symbols) of carbon black suspensions for three different preshear histories, and steady-state flow curve (crosses); lines: fit of the data to Herschel-Bulkley laws. Bottom: Start up flow at low shear ($\dot{\gamma} = 10^{-2}$ s$^{-1}$) (static yield stress measurements) after three different preshear histories. In both figures, the stress applied on the material during the preshear is 80.9 Pa (circles), 35.9 Pa (squares), and 24.7 Pa (triangles).*

For all studied preshear stress values $\tau_0$, the intrinsic behavior of the presheared material seems to be that of a classical simple yield stress fluid, with a plateau at low shear rates and a progressive increase of the slope (in semi-logarithmic scale) of the stress vs. shear rate with increasing shear rate. The yield stress value, associated with the level of the plateau, corresponds to the material dynamic yield stress of the material, and is a decreasing function of the preshear level $\tau_0$. Such a behavior contrasts with that of thixotropic materials: in that case the effective behavior of the material after preshear can be that of a viscous material with no yield stress[37,38,39], and a yield stress emerges due to the progressive restructuration of the material[15,35,36,39].

Let us now go one step further in the characterization of the solid behavior of the material obtained by the preshear. After preshear at a stress $\tau_0$, the material is left at rest at $\tau = 0$ Pa. Low amplitude oscillatory measurements then show that the material has a solid viscoelastic behavior (as already evidenced[25,26]). The material is subsequently sheared at a low constant shear rate $\dot{\gamma} = 10^{-2}$ s$^{-1}$. The results of the experiments are shown in Fig. 4b. The observed behavior is typical of the elasto-plastic behavior of yield stress fluids: the stress first increases basically linearly with the strain, which corresponds to the elastic straining of the material in its solid regime; at a given yield strain (of order 0.1), the shear stress reaches a plastic plateau, which defines the static yield stress of the material. It is worth noting that this static yield stress is equal to the dynamic yield stress obtained in the "intrinsic" flow curves of Fig. 4a. This is fully consistent with the description of the presheared material state as being that of a simple yield stress fluid.

A last important observation is that this static yield stress is independent of the time the material has been left at rest after preshear. This means that the structure does not evolve otherwise than under flow. This aspect strongly contrasts with thixotropic suspensions which basically restructure at rest.

In order to check the validity of these data at a local scale we have carried out tests with our MRI set up, where the velocity profiles are measured in time inside a Couette geometry. With this set up, we can only impose the rotation velocity of the inner cylinder. So our procedure consists in preshearing the material at a high rotation velocity (180 rpm) and then suddenly applying a lower velocity (from 80 to 2rpm), thus mimicking the rheometrical tests above based on stress decrease. Here we shall not enter into the detailed behavior; we only present the most important features, in relation with our macroscopic observations. From the local shear rate $\dot{\gamma}(R,t)$ (Eq. 1) and the local shear stress $\tau(R,t)$ (Eq. 2) we reconstruct the local flow behavior in two different situations. First, by collecting the ($\tau(R,t=0^+)$, $\dot{\gamma}(R,t=0^+)$) data measured at all radial positions $R$ just after the jump from a velocity $\Omega_0$ (here 180 rpm) to various other velocities $\Omega_i$ (here 40, 10, and 2 rpm), we get the "intrinsic" flow curve $\tau(\dot{\gamma})$ characteristic of the material presheared at $\Omega_0$. Then, by collecting ($\tau(R,t\to\infty)$, $\dot{\gamma}(R,t\to\infty)$) data in steady state at all radial positions $R$ and for various imposed rotational velocities, we get the steady-state flow curve $\tau(\dot{\gamma})$ (note that steady-state was studied here only at 180 and 2 rpm).

These local observations validate the macroscopic observations. Indeed we observe that the material is initially characterized by a simple a yield stress fluid flow curve (Fig. 5). We then observe that the material is rheopectic: the local flow curve observed at 2 rpm in Fig. 5 evolves towards larger shear rates and lower stresses (i.e., the local apparent viscosity decreases in time when the rotation velocity is lowered) until steady-state is reached. The observed local steady-state flow curve is in fair agreement with the macroscopic one. These results also show that the shear-induced softening under controlled rate can be spectacular: whereas the material has initially an intrinsic yield stress of order 30 Pa, it finally reaches stress values as low as 6 Pa at the lowest applied velocity. This shows that the material yield stress tends to disappear at low applied shear rate.

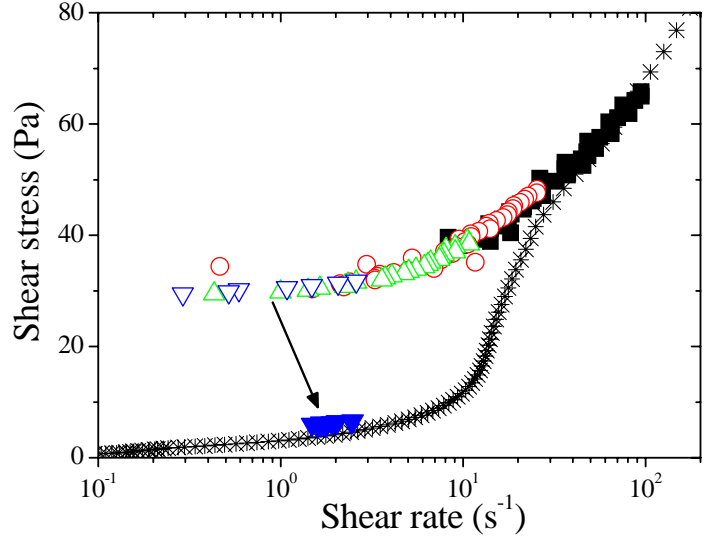

*Fig. 5*: *Intrinsic flow curve of the material presheared at 180 rpm (empty symbols) and steady-state flow curve (filled symbols), extracted from MRI measurements in a wide gap Couette geometry. Each symbol shape corresponds to measurements performed at a given rotational velocity $\Omega_i$: 180 rpm (squares), 40 rpm (circles), 10 rpm (up triangles), and 2 rpm (down triangles). The crosses are the steady-state flow curve measured in the macroscopic experiments.*

To summarize, the overall picture is puzzling as it does not correspond to the current knowledge on yield stress fluids. The main features of the material behavior, somehow contradictory, are the following: (i) there is no yield stress in the steady-state behavior; (ii) after a preshear, the materials shows viscosity bifurcation as thixotropic materials, but with history dependent $\tau_y$ and $\dot{\gamma}_c$; (iii) the intrinsic behavior of the presheared material is similar to that of a simple yield stress fluid, although this behavior is here unstable in contrast with these last materials.

In the following we try to describe the behavior of rheopectic yield stress fluids with the minimum ingredients. We show that all their original characteristics emerge naturally from the assumption that the material is more and more jammed as shear intensity is increased.

## 4. Theory

### *Basic modeling*

In order to model the behavior of rheopectic yield stress fluids and in particular better understand their specificities with regards to other materials, we first need to provide a description as simple as possible of all kinds of yield stress fluids in a single framework. As a starting point, we consider a material made of elements immersed in a liquid and possibly forming a jammed structure of the elements. In that case the simplest approach consists in writing the stress ($\tau$) as the sum of a contribution ($\lambda$) of the interactions between the elements and a contribution of the liquid phase flow[40]. This leads to a constitutive equation in simple shear of the form:

$\dot{\gamma} = 0$ if $\tau < \lambda$ (3a)

$\tau = \lambda + f(\dot{\gamma})$ if $\tau > \lambda$ (3b)

in which $\dot{\gamma}$ is the shear rate and $f(\dot{\gamma})$ the contribution to stress from the liquid flow. A further simplification consists in assuming that $f$ does not significantly depend on the evolutions of the structure, i.e. on the interactions between the elements. Under these conditions $f$ is continuous, increases with $\dot\gamma$ and is such that $f(0)=0$. On the contrary the stress contribution ($\lambda$) associated with element interactions depends on the current structure. More precisely, according to the current knowledge of the structure of such systems (see Sec. 2) this stress contribution can be associated with the critical stress value needed to break the network of interactions between the aggregates. For the sake of simplicity in the present context we will call $\lambda$ the degree of jamming.

Such a general model can be used to describe the behavior of both simple and thixotropic yield stress fluids. For a simple yield stress fluid, the degree of jamming is constant and $f$ is a power-law function of $\dot\gamma$. For a thixotropic yield stress fluids, the degree of jamming $\lambda$, i.e. the apparent yield stress of the material, depends on flow history. It decreases with the shear intensity. At rest, it increases with time.

In this framework, a rheopectic material should a priori be described as a material for which the degree of jamming $\lambda$ *increases* with shear intensity. In the following we examine the consequences of such a trend strictly within the frame of the simple above model.

From a general point of view $\lambda$ depends on the flow history, which may for example be expressed as $\lambda(t_0) = H\left[\tau(t)_{t<t_0}\right]$ in which $H$ is a functional. The advantage of using the stress for describing the flow history, instead of the shear rate, is that in systems developing flow heterogeneities this is a better controlled variable. We first focus on the response of such materials to creep flows. We assume that the material is initially in an unknown state of structure $\lambda(0)$. Let us now apply a constant preshear stress $\tau_1$ to the material over a very long time. If $\tau_1$ is sufficiently high (i.e., from Eq. 3, if $\tau_1 > \lambda(0)$), the material flows until reaching a steady state. As the functional $H$ only depends on $\tau_1$, we may write it in the following simpler form: $\lambda(t \to \infty) = F(\tau_1)$. For a rheopectic material, the degree of jamming should increase with the previous shear intensity: $F$ must then be an increasing function of $\tau_1$ and, since a steady-state flow in the liquid regime occurred for $\tau_1$ we have $F(\tau_1) < \tau_1$.

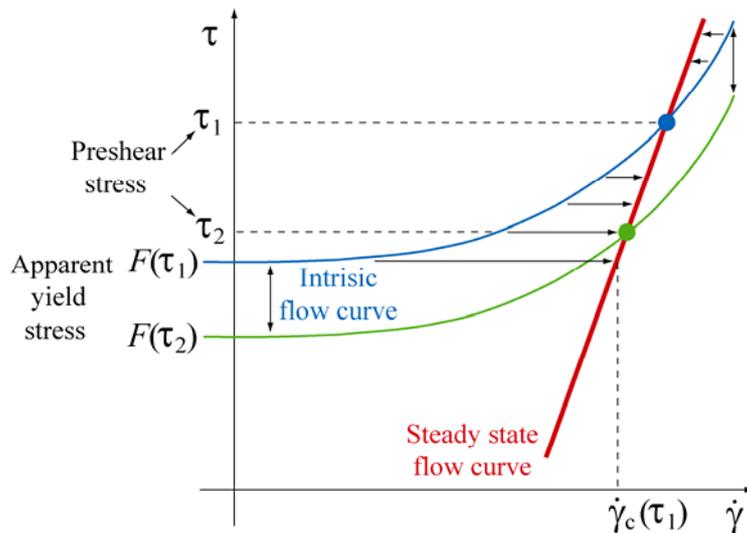

**Fig. 6**: *Instantaneous flow curve obtained under controlled stress just after steady flow at a given preshear stress $\tau_1$ (upper curve) and $\tau_2 < \tau_1$ (lower curve), and subsequent evolution of the shear rate in time towards steady state.*

*Response to creep flows*

Provided it is always possible to enforce flow at a stress $\tau_1$, this preshear leads to a reproducible initial state. Starting from this state, we now try to understand the response of the material to the simplest shear history, namely a creep flow. We thus consider a step change in the shear stress from $\tau_1$ to a new $\tau_2$ value.

At the end of the preshear, just before applying the new stress value, the material is in the state of structure $\lambda_1 = F(\tau_1)$. From Eq. 3, the material is thus initially a yield stress fluid with a yield stress $F(\tau_1)$, and an intrinsic flow behavior $\tau = F(\tau_1) + f(\dot{\gamma})$. If $\tau_2 < F(\tau_1)$, there is no flow: the system remains at rest (we shall not attempt to model here slow creep flows in the solid regime). If $\tau_2 > F(\tau_1)$, the material flows initially at a shear rate $\dot{\gamma}_2(t=0) = f^{-1}(\tau_2 - F(\tau_1))$. It is worth noting that, by varying $\tau_2$, the full set of ($\tau_2, \dot{\gamma}_2(t=0)$) values would completely characterize the intrinsic behavior of the presheared material, and describe the effective flow curve at the initial time $\tau = F(\tau_1) + f(\dot{\gamma})$, which is a simple yield stress fluid behavior. This response at short time is similar to what is observed experimentally (Fig. 4a).

Let us now look at the further evolution of the system for $\tau_2$ larger than the yield stress. If $\tau_2 < \tau_1$, the shear rate $\dot{\gamma}_2(t)$ will progressively reach the steady-state associated to a steady shear at $\tau_2$, characteristic of a structural parameter $\lambda_2 = F(\tau_2) < \lambda_1$; i.e., from Eq. 3, we have $\dot{\gamma}_2(t=\infty) = f^{-1}(\tau_2 - F(\tau_2))$. Given the properties of $f$ and $F$, we note that $\dot{\gamma}_2(t=\infty) > \dot{\gamma}_2(t=0)$, i.e., when the shear stress is lowered after the preshear, the shear rate increases during the creep flow and the degree of jamming decreases. An opposite effect occurs if a stress $\tau_2$ larger than $\tau_1$ is imposed: now the shear rate decreases to its steady state value (see Fig. 6) while the degree of jamming increases. As a consequence we are dealing with a *rheopectic* material; this behavior contrasts with that of thixotropic materials, for which the shear rate decreases when the applied stress is lowered, and increases when the applied stress is increased.

Let us now look at the steady shear rate values $\dot{\gamma}_2(t=\infty) = f^{-1}(\tau_2 - F(\tau_2))$ reached for such flow histories. As flow occurs only if $\tau_2 > F(\tau_1)$, this implies that all steady state shear rates are higher than a critical value $\dot{\gamma}_c(\tau_1) = f^{-1}(F(\tau_1) - F(F(\tau_1)))$. We thus recover the viscosity bifurcation phenomenon observed experimentally. After a preparation at a stress $\tau_1$ the material will not be able to flow in steady state at a shear rate smaller than a finite, critical value $\dot{\gamma}_c(\tau_1)$. Such a characteristic is typically observed for thixotropic materials with a degree of jamming decreasing with the preshear intensity but not for simple yield stress fluids[9,11,15,39,41]. It is remarkable that this characteristics is also observed with materials with an opposite variation of the degree of jamming with preshear intensity. However, in contrast with thixotropic fluids, here this does not lead to a shear-banding effect under controlled shear rate: if a shear rate is imposed below the critical shear rate the material will a priori reach a homogenous flow at the shear stress corresponding to this shear rate along the flow curve given by Eq. 4 (as observed experimentally, see Fig. 5).

Moreover, although similar behavior is predicted for any value of the preshear stress $\tau_1$, we note that the mechanical characteristics of viscosity bifurcation depend on the value of $\tau_1$ since the yield stress is $\tau_y(\tau_1) = F(\tau_1)$ and the critical shear rate is $\dot{\gamma}_c(\tau_1) = f^{-1}(F(\tau_1) - F(F(\tau_1)))$. These features, also observed experimentally (Fig. 1), appear naturally as a specificity of rheopectic yield stress materials.

From the above results we deduce that the complete steady state flow curve should express as:
$$\dot{\gamma} = f^{-1}(\tau - F(\tau)) \tag{4}$$
We emphasize that when creep flows are imposed after a preshear at a stress $\tau_1$, only the part of the curve corresponding to $\dot{\gamma} > \dot{\gamma}_c(\tau_1)$ is obtained, as observed experimentally (Fig. 3b). Actually one must keep in mind that the full steady-state flow curve can be described only for an appropriate flow history including a progressive stress decrease.

To summarize, from the simplest possible equation for yield stress fluids (Eq. 3), by assuming that the degree of jamming increases with shear intensity, we recover the most important features observed experimentally: a rheopectic behavior, and a viscosity bifurcation with history dependent yield stress and critical shear rate (of value decreasing with the value of the preshear stress). It is worth emphasizing that these original results rely on a very limited number of assumptions at the basis of the model, so that they can be considered as general consequences of this behavior.

In order to go one step further in the comparison with the experiments we now need to make hypotheses on the functions and to describe explicitly the time evolution of the structural parameter.

*Additional modeling hypotheses*

We now assume that $F(\tau) < \tau$ for any $\tau > 0$ and that $F(0) = 0$. From the steady-state flow curve (Eq. 4), this simple additional hypothesis implies that a flow in the liquid regime can a priori be obtained under any stress value, as observed experimentally, although viscosity bifurcation is predicted for a series of creep flows. Let us show how it is possible to get such a vanishing yield stress. After a preparation at a stress $\tau_1$ the apparent yield stress of the material is $F(\tau_1)$. This apparent yield stress can be lowered if the material is prepared at a smaller stress $\tau_2 < \tau_1$, provided it flows when $\tau_2$ is applied. In other words, if we start from the material preparation at $\tau_1$, $\tau_2$ needs to be larger than the yield stress reached after the initial preparation, i.e., $\tau_2 > F(\tau_1)$. Finally, with such successive stress steps of decreasing values larger than the apparent yield stress reached during the step before, we can progressively decrease the degree of jamming of the material, i.e., its yield stress. It decreases down to zero under the hypothesis we have made on $F$. This means that, in contrast with other kinds of yield stress fluid, such a material should be able to flow steadily under any stress level.

Full modeling requires a kinetic equation for the structural parameter. Let us now attempt to describe the way the degree of jamming reaches the steady state. The degree of jamming represents the force per unit surface required to separate some fraction of links existing between the aggregates. As a consequence it can be associated with some level of energy in the system: roughly speaking $\lambda$ is the sum of the potential energy of these links per unit volume. Depending on the material history, these links will be either destroyed or created during flow when a stress $\tau$ is applied; if $\tau$ is applied for a sufficient time the system will reach a steady degree of jamming, or level of energy, $F(\tau)$. At the scale of an aggregate we can see the process as a repeated attempt to reach the level of energy $F(\tau)$. If the current level of energy is $\lambda$, an elementary change of $\lambda$, i.e. $d\lambda$, during some period of time $dt$, is proportional to the number of attempts of link formation or breakage, and to some function of the energy barrier $F(\tau) - \lambda$ to overcome. The simplest way for representing the latter function consists in assuming that it is proportional to this barrier; the number of attempts is simply proportional to the relative motion of the aggregates and thus to the deformation undergone by the sample, i.e., $\dot{\gamma} dt$. Finally we get: $d\lambda/dt = a\dot{\gamma}(F(\tau) - \lambda)$, in which $a$ is a material parameter. In addition we assume that $\lambda$ remains constant if the material is at rest ($\dot{\gamma} = 0$). This is consistent with the general kinetic equation above. As a consequence there is a possible evolution of the degree of jamming only during flow.

Finally the complete model expresses as follows:
$\dot{\gamma} = 0$ if $\tau < \lambda$
$\tau = \lambda + f(\dot{\gamma})$ if $\tau > \lambda$
$$\frac{d\lambda}{dt} = a\dot{\gamma}(F(\tau) - \lambda) \tag{5}$$
The model is thus entirely defined when $F$, $f$ and $a$ have been determined.

*Comparison theory-experiments*

As shown above, from a qualitative viewpoint the experimental results correspond exactly to the predictions of the theoretical model proposed above on the basis of extremely simple assumptions. We can further check the consistency of the rheological behavior of this material with the predictions of the model.

We start by looking at the intrinsic flow curves obtained after different preshear levels. If we subtract the yield stress from each of these flow curves we should obtain the function $f$ accounting for viscous dissipation. The consistency of our data with regards to the model is proved by the fact that the intrinsic flow curves obtained for different preshear stress levels are similar by a simple vertical translation (see Fig. 4a), which means that $\tau - \tau_y$ is independent of the preshear level as assumed by the model.

In order to make a more complete comparison of the data with the model we have applied the following procedure. The material is always first presheared at a given value $\tau_p = 80.9$ Pa. The stress value is then decreased step by step (as explained above, to ensure that flow never stops) towards a test stress $\tau_i$; the material is then sheared during 1 h at $\tau_i$ and the corresponding steady-state shear rate $\dot{\gamma}_i = \dot{\gamma}(\tau_i, t \to \infty)$ is measured. From this state, two different experiments are performed: (i) the material is left at rest and a yield stress measurement at low imposed shear rate is performed (as in Fig. 4b), from which the state dependent yield stress $\tau_y(\tau_i)$ is obtained; (ii) the shear stress is abruptly changed to $\tau_p$, and the maximum shear rate $\dot{\gamma}_p = \dot{\gamma}(\tau_p, t = 0^+)$ measured just after the step change in stress is considered to characterize the intrinsic behavior of the material sheared at a stress $\tau_i$. If we assume that the intrinsic flow curve is $\tau = \tau_y(\tau_i) + k(\tau_i)\dot{\gamma}^{n(\tau_i)}$, all parameters are determined from these experiments. The structure dependent yield stresses have been shown in Fig. 2. The exponents $n(\tau_i)$, computed as $[\log(\tau_p - \tau_y(\tau_i)) - \log(\tau_i - \tau_y(\tau_i))] / [\log(\dot{\gamma}_p) - \log(\dot{\gamma}_i)]$, are depicted in Fig. 7. It is seen that $n$ remains close to 0.5 for preshear stresses larger than 25 Pa, and increases slightly below this value. To compare the $k(\tau_i)$ values obtained in all experiments, a value of $n = 0.5$ has been assumed in all cases, and $k(\tau_i)$ was computed as $k = (\tau_p - \tau_i) / (\dot{\gamma}_p^{0.5} - \dot{\gamma}_i^{0.5})$ ; these values are shown in Fig. 7 : k seems to remain constant in the whole range of preshear stresses, in agreement with the assumption of the model, i.e. viscous dissipation does not depend on the structure.

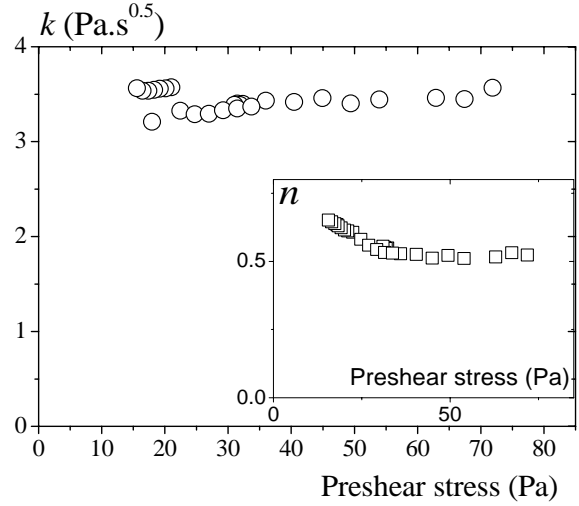

*Fig. 7*: *Parameters n and k of the fit of the intrinsic flow curve to a Herschel-Bulkley model as a function of the preshear stress (see exact procedure in text).*

An additional inspection of the consistency of the model can be carried out. The apparent yield stresses measured after different preshear levels directly give us the function $F(\tau)$; we have also determined experimentally the steady-state flow curve. We can thus deduce the function $f$ from these data and Eq. 4. This result can be compared to the function $f$ obtained from the intrinsic flow curves of Fig. 4a. We find a reasonable agreement between the results obtained for $f$ from these two sides (see Fig. 8), which confirms the consistency of the data with regards to the model.

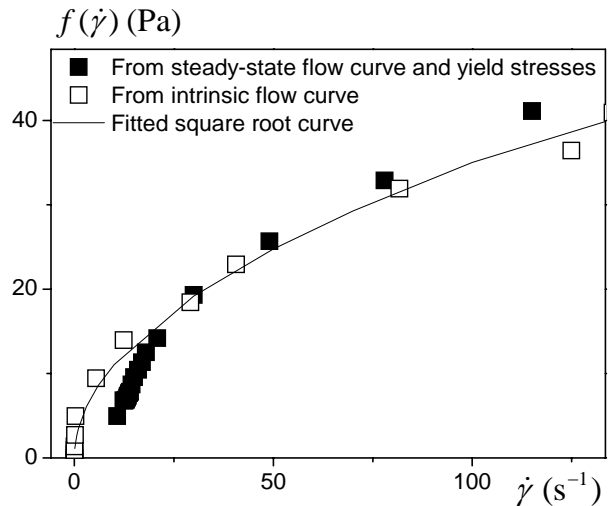

*Fig. 8*: *Function $f(\dot{\gamma})$ in the model (Eq. 3), assumed to be valid for a carbon black suspension, and obtained from (filled squares) measurements of the function $F$ and of the steady-state flow curve or (empty squares) from the instantaneous flow curve (after a preshear of 80.9 Pa). The line is a curve $k\dot{\gamma}^{0.5}$ with $k = 3.5$ $Pa.s^{0.5}$.*

Let us now compare the time evolutions predicted by the model with the experimental ones. We consider creep tests (at various values $\tau_2$) after a preshear at a given stress $\tau_1$. We see that the model is able to predict very well all the trends observed in a creep test series (see Fig. 9) with a single value for $a$ (i.e. 50): the theoretical and experimental curves have similar shapes and the characteristic time for which the rapid increase in shear rate occurs is well predicted by the theory whereas it varies over several decades when one closely approaches the yield stress. Note that although the model has been separately fitted to the intrinsic flow curve this has been done in a global way (i.e. with the whole set of data we had) and with various sources of uncertainty at different steps. This explains that there is not a strict agreement of the initial levels of the shear rate in this comparison for stresses close to the yield stress. However considering the uncertainty on the exact shear stress effectively applied on a sample in rheometrical tests the global agreement between the theory and the experimental data is quite satisfactory and demonstrates the relevance of the model.

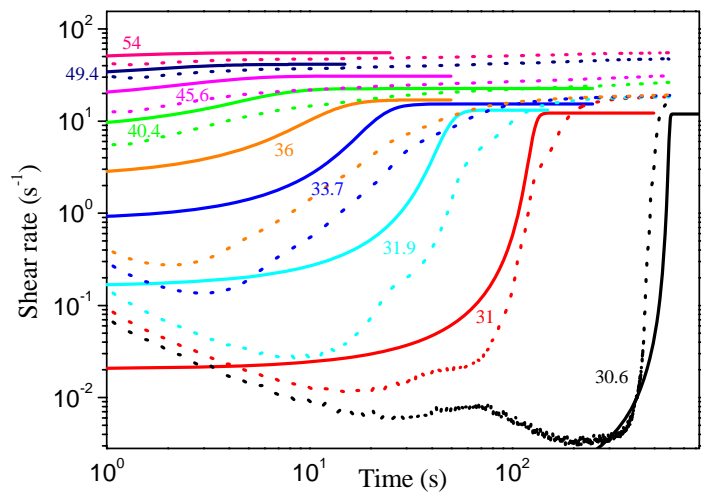

*Fig. 9*: *Comparison of theoretical predictions (continuous curves) of the model with experimental data (dotted lines) for a series of creep tests after a preshear at 80.9 Pa. The same colors have been used for similar stress values; these values are given in the figure.*

## 5. Conclusion

We have shown that a yielding rheopectic material exhibits original properties. After a preparation at a given stress its intrinsic flow curve is that of a yield stress fluid but then evolves with a viscosity bifurcation effect. This trend nevertheless differs from that observed for thixotropic fluids, since the yield stress and the critical shear rate now depend on the previous flow history. Moreover, here this effect does not imply that shear-banding will occur at a shear rate imposed below the critical shear rate. In fact the material is able to flow homogeneously under almost any conditions (i.e. at any shear rate or shear rate) as soon as it has been prepared appropriately. This is due to the fact that its structure can be adjusted at will through an appropriate flow history. In particular it is possible to suppress the yield effect via a very progressive stress decrease so that the material is finally in the state of a simple liquid. A spectacular point is that as soon as it has been prepared in such a way the material remains in the state of structure it has reached before stoppage. This critically contrasts with the behavior of usual thixotropic fluids which have a structure which strengthens at rest.

We have developed a very simple model relying on natural arguments, in particular based on the assumption that the structure evolves only as a result of flow. It appears that this model is able to capture quantitatively very well all the trends observed for these materials. This suggests that the rheological properties observed for our carbon black suspensions very general for rheopectic yield stress fluids.